

\input lanlmac

\input epsf

\newcount\figno
\figno=0
\def\fig#1#2#3{
\par\begingroup\parindent=0pt\leftskip=1cm\rightskip=1cm\parindent=0pt
\baselineskip=11pt
\global\advance\figno by 1
\midinsert
\epsfxsize=#3
\centerline{\epsfbox{#2}}
\vskip 12pt
{\bf Fig. \the\figno ~~} #1\par
\endinsert\endgroup\par
}
\def\figlabel#1{\xdef#1{\the\figno}}


\def\CS{{\cal S}}
\def\th{\theta}

\def\S{{\bf S}}
\def\Z{{\bf Z}}

\def\Tr{{\rm Tr}}
\def\hf{{1\over 2}}
\def\qu{{1\over 4}}
\def\R{{\bf R}}
\def\o{\over}
\def\til#1{\widetilde{#1}}
\def\si{\sigma}

\def\b#1{\overline{#1}}
\def\del{\partial}
\def\wg{\wedge}
\def\lap{\Delta}
\def\bra{\langle}
\def\ket{\rangle}
\def\lf{\left}
\def\ri{\right}
\def\riya{\rightarrow}

\def\lrya{\leftrightarrow}

\def\J{{\cal J}}
\def\la{\lambda}

\def\h#1{\widehat{#1}}

\def\bt{\beta}

\def\Ga{\Gamma}
\def\al{\alpha}

\def\dag{\dagger}
\def\rt#1{\sqrt{#1}}

\def\sitarel#1#2{\mathrel{\mathop{\kern0pt #1}\limits_{#2}}}

\def\cob{\delta}

\lref\LMSp{
H.~Liu, G.~Moore and N.~Seiberg,
``Strings in a time-dependent orbifold,''
JHEP {\bf 0206}, 045 (2002)
[hep-th/0204168].
}
\lref\LMSn{
H.~Liu, G.~Moore and N.~Seiberg,
``Strings in time-dependent orbifolds,''
JHEP {\bf 0210}, 031 (2002)
[hep-th/0206182].
}
\lref\Mc{
M.~Fabinger and J.~McGreevy,
``On smooth time-dependent orbifolds and null singularities,''
[hep-th/0206196].
}
\lref\HoroPol{
G.~T.~Horowitz and J.~Polchinski,
``Instability of spacelike and null orbifold singularities,''
[hep-th/0206228].
}
\lref\Albion{
A.~Lawrence,
``On the instability of 3D null singularities,''
[hep-th/0205288].
}

\lref\Bachas{
C.~Bachas and C.~Hull,
``Null brane intersections,''
[hep-th/0210269].
}
\lref\Myers{
R.~C.~Myers and D.~J.~Winters,
``From D - anti-D pairs to branes in motion,''
[hep-th/0211042].
}

\lref\Figue{
J.~Figueroa-O'Farrill and J.~Simon,
``Generalized supersymmetric fluxbranes,''
JHEP {\bf 0112}, 011 (2001)
[hep-th/0110170].
}
\lref\Simon{
J.~Simon,
``The geometry of null rotation identifications,''
JHEP {\bf 0206}, 001 (2002)
[hep-th/0203201].
}
\lref\FSone{
J.~Figueroa-O'Farrill and J.~Simon,
``Supersymmetric Kaluza-Klein reductions of M2 and M5 branes,''
[hep-th/0208107].
}
\lref\FStwo{
J.~Figueroa-O'Farrill and J.~Simon,
``Supersymmetric Kaluza-Klein reductions of M-waves and MKK-monopoles,''
[hep-th/0208108].
}

\lref\HoroSteif{
G.~T.~Horowitz and A.~R.~Steif,
``Singular String Solutions With Nonsingular Initial Data,''
Phys.\ Lett.\ B {\bf 258}, 91 (1991).
}
\lref\Nekrasov{
N.~A.~Nekrasov,
``Milne universe, tachyons, and quantum group,''
[hep-th/0203112].
}

\lref\Sethi{
A.~Hashimoto and S.~Sethi,
``Holography and string dynamics in time-dependent backgrounds,''
[hep-th/0208126].
}
\lref\Alisha{
M.~Alishahiha and S.~Parvizi,
``Branes in time-dependent backgrounds and AdS/CFT correspondence,''
JHEP {\bf 0210}, 047 (2002)
[hep-th/0208187].
}
\lref\Ohta{
R.~G.~Cai, J.~X.~Lu and N.~Ohta,
``NCOS and D-branes in time-dependent backgrounds,''
[hep-th/0210206].
}
\lref\Nappi{
L.~Dolan and C.~R.~Nappi,
``Noncommutativity in a time-dependent background,''
[hep-th/0210030].
}

\lref\Kh{
J.~Khoury, B.~A.~Ovrut, N.~Seiberg, P.~J.~Steinhardt and N.~Turok,
``From big crunch to big bang,''
Phys.\ Rev.\ D {\bf 65}, 086007 (2002)
[hep-th/0108187].
}
\lref\Elitzur{
S.~Elitzur, A.~Giveon, D.~Kutasov and E.~Rabinovici,
``From big bang to big crunch and beyond,''
JHEP {\bf 0206}, 017 (2002)
[hep-th/0204189].
}
\lref\Cornalb{
L.~Cornalba, M.~S.~Costa and C.~Kounnas,
``A resolution of the cosmological singularity with orientifolds,''
Nucl.\ Phys.\ B {\bf 637}, 378 (2002)
[hep-th/0204261].
}
\lref\Cornala{
L.~Cornalba and M.~S.~Costa,
``A New Cosmological Scenario in String Theory,''
Phys.\ Rev.\ D {\bf 66}, 066001 (2002)
[hep-th/0203031].
}

\lref\Craps{
B.~Craps, D.~Kutasov and G.~Rajesh,
``String propagation in the presence of cosmological singularities,''
JHEP {\bf 0206}, 053 (2002)
[hep-th/0205101].
}
\lref\Dudas{
E.~Dudas, J.~Mourad and C.~Timirgaziu,
``Time and space dependent backgrounds from nonsupersymmetric strings,''
[hep-th/0209176].
}

\lref\bath{
O.~Aharony, M.~Fabinger, G.~T.~Horowitz and E.~Silverstein,
``Clean time-dependent string backgrounds from bubble baths,''
JHEP {\bf 0207}, 007 (2002)
[hep-th/0204158].
}
\lref\nothing{
V.~Balasubramanian and S.~F.~Ross,
``The dual of nothing,''
Phys.\ Rev.\ D {\bf 66}, 086002 (2002)
[hep-th/0205290].
}
\lref\Balas{
V.~Balasubramanian, S.~F.~Hassan, E.~Keski-Vakkuri and A.~Naqvi,
``A space-time orbifold: A toy model for a cosmological singularity,''
[hep-th/0202187].
}
\lref\Eva{
A.~Maloney, E.~Silverstein and A.~Strominger,
``De Sitter space in noncritical string theory,''
[hep-th/0205316].
}
\lref\Simonhol{
J.~Simon,
``Null orbifolds in AdS, time dependence and holography,''
JHEP {\bf 0210}, 036 (2002)
[hep-th/0208165].
}

\Title{             
                                             \vbox{\hbox{EFI-02-51}
                                             \hbox{hep-th/0211218}}}
{\vbox{
\centerline{D-Branes on the Null-Brane}
}}

\vskip .2in

\centerline{Kazumi Okuyama}

\vskip .2in

\centerline{ Enrico Fermi Institute, University of Chicago} 
\centerline{ 5640 S. Ellis Ave., Chicago IL 60637, USA}
\centerline{\tt kazumi@theory.uchicago.edu}

\vskip 3cm
\noindent

We study D-branes in the null-brane background.
Using the covariant formalism of the worldsheet theory,
we construct the boundary states describing D-branes on the null-brane.
From the cylinder amplitudes,
we find that the D-branes with codimension zero or two
have time-dependent effective tensions.

\Date{November 2002}

\vfill
\vfill

\newsec{Introduction}
It is important to study string theory in various curved backgrounds
since our understanding of string theory is not background
independent.
In particular, it is interesting to study string theory in
time-dependent backgrounds.
Time-dependent orbifolds of flat space
would be good toy models for more realistic
cosmological backgrounds in string theory \refs{\HoroSteif
\Kh\Figue\Balas\Cornala\Nekrasov\Simon\LMSp\bath\Elitzur\Cornalb\Craps
\Albion\nothing\Eva\LMSn\Mc\HoroPol\Sethi\Simonhol{--}\Dudas}.

The null-brane \refs{\Figue,\Simon} is an orbifold of $\R^{1,3}$
which has several good properties as a background of string theory:
It is a smooth supersymmetric
orbifold and does not contain  closed time-like curves.
It was shown that the gravitational backreaction 
due to the introduction of a single particle in this background
is supressed when the number of noncompact dimensions 
is larger than five \HoroPol, and that the tree level string amplitude
makes sence in this background \refs{\LMSn,\Mc}.


It is interesting to study the behavior of solitonic objects
in time-dependent backgrounds.
In this paper, we consider D-branes in the null-brane background
and construct the boundary states describing them.
After reviewing the geometry of the null-brane in section 1, 
in section 2 we consider the closed string theory on the null-brane.
We find that the Virasoro operator in the twisted sector is related
to the one in the untwisted sector by a spectral flow.
Using the operator formalism, 
we rederive the torus amplitude obtained in \refs{\LMSn,\Mc}.
In section 3, we construct the boundary states for D3- and D2-branes,
and compute the cylinder amplitudes.
We find that the effective tension of the D3-brane is time-dependent.
In section 4, we consider D1-brane and discuss its relation to 
the null-scissors configuration discussed recently in \Bachas.
Section 5 is devoted to discussions. 
In appendix A, we consider the trace in the zero-mode part of a string,
and in appendix B we discuss the T-dual of the null-scissors.

\newsec{Geometry of the Null-Brane}
The null-brane is an orbifold of a flat 4-dimensional
Minkowski space $\R^{1,2}\times\R$ with the metric
\eqn\metflat{
ds^2=\eta_{\mu\nu}dx^\mu dx^\nu+dz^2=-2dx^+dx^-+dx^2+dz^2.
}
The orbifold group $\Ga\simeq\Z$ 
acts as a null-boost on $x^\mu(\mu=+,x,-)$ and shifts
$z$ by a constant:
\eqn\orbfld{
X=\lf(\matrix{x^+\cr x\cr x^-}\ri)\mapsto e^{2\pi\J}X,\quad
z\mapsto z+2\pi R.
}
Above, the matrix $\J$ is the vector representation of
the generator $J=(J^{0x}+J^{1x})/\rt{2}$, and its explicit form
is given by
\eqn\Jmat{
\J=\lf(\matrix{0&0&0\cr1&0&0\cr0&1&0}\ri).
}
The important property
of $\J$ is that it is nilpotent 
and belongs to the Lie algebra $so(1,2)$
\eqn\Jrel{
\J^3=0,\quad \J^T=-\eta \J\eta^{-1},
}
where the superscript $T$ denotes the transpose of the matrix.

To describe the geometry of the orbifold $(\R^{1,2}\times\R)/\Ga$,
it is convenient to introduce a new set of coordinates
$(y^+,y,y^-,u)$ defined as
\eqn\yudef{
\lf(\matrix{x^+\cr x\cr x^-}\ri)
=e^{y\J}\lf(\matrix{y^+\cr 0\cr y^-}\ri)
=\lf(\matrix{y^+\cr yy^+\cr y^-+\hf y^2y^+}\ri),\quad
z=u+Ry.
}
In terms of these coordinates, the orbifold action \orbfld\ becomes 
simply $y\sim y+2\pi$ 
with other coordinates kept invariant.
The metric on the null-brane in these variables becomes
\eqn\metnulyu{
ds^2=-2dy^+dy^-+[R^2+(y^+)^2](dy+A)^2+{(y^+)^2\o R^2+(y^+)^2}du^2
}
where $A$ is the connection of the $\S^1$ fiber
with non-vanishing curvature
\eqn\AFinyu{
A={R\o R^2+(y^+)^2}du,\quad F=dA=-{Ry^+\o R^2+(y^+)^2}dy^+\wg du.
}

\newsec{Closed Strings on the Null-Brane}
\subsec{Mode Expansion in the Twisted Sector}
To construct the boundary states on the null-brane, let us begin with 
the mode expansion of worldsheet fields in the covariant formalism.
First we focus on the $\R^{1,2}$ part of the target space
whose worldsheet action is
\eqn\wsS{
S={1\o4\pi\al'}\int dt\int_0^{2\pi}d\si(\del_tX^\mu\del_t X_{\mu}
-\del_\si X^\mu\del_\si X_{\mu})
={1\o\pi\al'}\int dt\int_0^{2\pi}d\si\del_+X^T\eta \del_-X
}
where 
\eqn\upm{
u^{\pm}=t\pm\si,\quad
\del_{\pm}\equiv {\del\o\del u^{\pm}}=\hf(\del_t\pm\del_\si).
}

In the $w^{\rm th}$ twisted sector, $X(t,\si)$ 
is periodic up to
a $\Ga$-action
\eqn\Xperi{
X(t,\si+2\pi)=e^{2\pi w\J}X(t,\si).
}
The mode expansion of $X$ in this sector is given by \LMSp
\eqn\modeX{
X=i\rt{\al'\o2}\sum_{n\in\Z}\lf({e^{-i(n+iw\J)u^+}\o n+iw\J}\al_n^{(w)}
+{e^{-i(n-iw\J)u^-}\o n-iw\J}\til{\al}_n^{(w)}\ri),
}
and its conjugate momentum is
\eqn\momPi{
\Pi={\del_tX\o2\pi\al'}={1\o2\pi\rt{2\al'}}\sum_{n\in\Z}
\Big(e^{-i(n+iw\J)u^+}\al_n^{(w)}+e^{-i(n-iw\J)u^-}\til{\al}_n^{(w)}\Big).
}
The zero modes are written as
\eqn\alzero{
\al_0^{(w)}=\rt{\al'\o2}(p+{w\J\o\al'}x),\quad
\til{\al}_0^{(w)}=\rt{\al'\o2}(p-{w\J\o\al'}x),
}
where  $x^\mu$ and $p^\mu$ satisfy
the canonical commutation relation $[x^\mu,p^\nu]=i\eta^{\mu\nu}$.
At $t=0$, \modeX\ and \momPi\  reduce to
\eqn\XPitzero{\eqalign{
X&=e^{w\si\J}\lf[x+i\rt{\al'\o2}
\sum_{n\not=0}{e^{-in\si}\o n+iw\J}(\al_n^{(w)}-\til{\al}^{(w)}_{-n})
\ri]\equiv e^{w\si\J}\til{X} \cr
\Pi&={1\o2\pi}e^{w\si\J}\lf[p+{1\o\rt{2\al'}}
\sum_{n\not=0}e^{-in\si}(\al_n^{(w)}+\til{\al}_{-n}^{(w)})\ri]\equiv
e^{w\si\J}\til{\Pi}.
}}


From the relation $[X^\mu(\si),\Pi_\nu(\si')]
=i\cob^\mu_\nu\cob(\si,\si')$,
the commutation relations of the non-zero modes
are found to be 
\eqn\commodes{\eqalign{
[\al_n^{(w)\mu},\al_m^{(w)\nu}]
&=(n\eta^{-1}+iw\J\eta^{-1})^{\mu\nu}\cob_{n+m,0}~~,\cr
[\til{\al}_n^{(w)\mu},\til{\al}_m^{(w)\nu}]
&=(n\eta^{-1}-iw\J\eta^{-1})^{\mu\nu}\cob_{n+m,0}~~.
}}
More explicitly, the non-vanishing commutators for the left movers are
\eqn\comalc{
[\al_n^{(w)x},\al_m^{(w)x}]=n\cob_{n+m,0},\quad
[\al_n^{(w)+},\al_m^{(w)-}]=-n\cob_{n+m,0},\quad
[\al_n^{(w)-},\al_m^{(w)x}]=iw\cob_{n+m,0}.
}

Next consider the field $Z(t,\si)$ in the twisted sector
$Z(t,\si+2\pi)=Z(t,\si)+2\pi wR$.
The mode expansion of $Z$ is the familiar winding configuration on a circle:
\eqn\modeZ{
Z=z+\rt{\al'\o2}(\al_0^{(w)z}u^++\til{\al}_0^{(w)z}u^-)+
i\rt{\al'\o2}\sum_{n\not=0}
{1\o n}(e^{-inu^+}\al_n^{z}+e^{-inu^-}\til{\al}_n^{z})
}
where $[\al_n^z,\al_m^z]=n\cob_{n+m,0}$ and
\eqn\alzeroz{
\al_0^{(w)z}=\rt{\al'\o2}p_z+{wR\o\rt{2\al'}},\quad
\til{\al}_0^{(w)z}=\rt{\al'\o2}p_z-{wR\o\rt{2\al'}}.
}

\subsec{Spectral Flow}
As pointed out in \LMSp, the commutation relation \commodes\ leads to
an exchange algebra of the currents $\del_{\pm}X^\mu$.
Therefore, one might think that the quantization of the twisted
sector would be quite complicated.
However, by a simple change of variables the relation \commodes\
reduces to the ordinary commutation relation of oscillators.  
To see this, we introduce the following invertible matrix $\CS_{\beta}$
\eqn\Sbtdef{
\CS_\bt=1+{i\o2}\bt\J+{1\o8}\bt^2\J^2,\quad
(\CS_\bt)^{-1}=1-{i\o2}\bt\J-{3\o8}\bt^2\J^2.
}
which satisfies
\eqn\relSbt{
(\CS_\bt)^2=1+i\bt\J,\quad (\CS_\bt)^T=\eta\CS_{-\bt}\eta^{-1}.
}
In terms of this, the commutation relation \commodes\ is rewritten as
\eqn\comosSbt{\eqalign{
[\al_n^{(w)\mu},\al_m^{(w)\nu}]&=n\cob_{n+m,0}
\Big(\CS_{w\o n}\eta^{-1}\CS_{w\o m}^T\Big)^{\mu\nu},\cr
[\til{\al}_n^{(w)\mu},\til{\al}_m^{(w)\nu}]&=n\cob_{n+m,0}
\Big(\CS_{-{w\o n}}\eta^{-1}\CS_{-{w\o m}}^T\Big)^{\mu\nu}.
}}
In other words, $\CS_{\pm{w\o n}}$ are the square root of the matrices
in the right-hand sides of \commodes.
Therefore, the oscillators in the twisted sector and the ordinary
untwisted oscillators are related by the following ``spectral flow''
\eqn\flowalw{
\al_n^{(w)}=\CS_{w\o n}\al_n^{(0)},\quad
\til{\al}_n^{(w)}=\CS_{-{w\o n}}\til{\al}_n^{(0)}.
}
Namely, by taking a linear combination
of the oscillators in the twisted sector, we can construct 
$\al^{(0)}_n$ and $\til{\al}^{(0)}_n$ obeying
the ordinary relation
\eqn\alzeroc{
[\al^{(0)\mu}_n,\al^{(0)\nu}_m]
=[\til{\al}^{(0)\mu}_n,\til{\al}^{(0)\nu}_m]=n\cob_{n+m,0}\eta^{\mu\nu}.
}
The oscillator vacuum in the $w^{\rm th}$ twisted sector 
is naturally defined as
\eqn\vac{
\al^{(w)\mu}_{n}|0\ket_w=\al_{n}^{(0)\mu}|0\ket_w=0\qquad(n>0).
}

\subsec{Symmetry Generators}
In this subsection, we construct the generator of the
null-boost and the Virasoro operator $L_0$ in the twisted sector.

Let us first consider the null-boost generator $\h{J}$:
\eqn\Jpx{
\h{J}=\int_0^{2\pi}d\si(X^+\Pi^x-X^x\Pi^+)=
-\int_0^{2\pi}d\si X^T\eta\J\Pi
=-\int_0^{2\pi}d\si \til{X}^T\eta\J\til{\Pi},
}
where the variables $\til{X}$ and $\til{\Pi}$ are defined in \XPitzero.
In the last equality we used the 
fact that $e^{w\si\J}$ preserves $\eta$.
Now $\h{J}$ is written in terms of the modes as 
\eqn\JinXtil{
\h{J}
=J_0+E+\til{E},
}
where
\eqn\JzeroE{\eqalign{
J_0&=-x^T\eta\J p=x^+p^x-xp^+ \cr
E&={i\o2}\sum_{n\not=0}{1\o n}\al_{-n}^{(0)T}\eta\J\al_{n}^{(0)}
=i\sum_{n\not=0}{1\o n}\al_n^{(0)+}\al_{-n}^{(0)x},\quad
\til{E}=i\sum_{n\not=0}{1\o n}\til{\al}_n^{(0)+}\til{\al}_{-n}^{(0)x}.
}}
Here we wrote $E$ and $\til{E}$ in terms of the untwisted oscillators.

Next consider the stress tensor in the twisted sector.
The $\R^{1,2}\times\R$ part of the stress tensor is given by
\eqn\EMT{
T={1\o\al'}\del_+X^\mu\del_+X_{\mu}
+{1\o\al'}(\del_+Z)^2\equiv\sum_{n\in\Z}e^{-in\si}L_n^{(w)}.
}
By substituting the mode expansion, $L_0$ in the 
${w}^{\rm th}$ twisted sector is found to be
\eqn\Lzerow{
L_0^{(w)}=\hf\sum_{n\in\Z}\al^{(w)\mu}_n\al_{-n,\mu}^{(w)}
+\hf\sum_{n\in\Z}\al^{z}_n\al_{-n}^{z}.
}
We can rewrite $L_0^{(w)}$ in terms of the untwisted oscillator
using the spectral flow \flowalw:
\eqn\Lzeroflow{\eqalign{
L_0^{(w)}&={\al'\o4}(p^\mu p_\mu+p_z^2)
+N^{(0)}+{w\o2}(J_0+p_zR+2E)+{\al'\o4}\lf({2\pi wR_+\o2\pi\al'}\ri)^2
\cr
\til{L}_0^{(w)}&={\al'\o4}(p^\mu p_\mu+p_z^2)
+\til{N}^{(0)}-{w\o2}(J_0+p_zR+2\til{E})
+{\al'\o4}\lf({2\pi wR_+\o2\pi\al'}\ri)^2
}}
where $N^{(0)}$ and $\til{N}^{(0)}$ are the number operators of the 
untwisted oscillators, and  
$R_+$ is defined as
\eqn\Rxp{
R_+=\rt{(x^+)^2+R^2}.
}
The last term in \Lzeroflow\ can be understood as the $({\rm mass})^2$
of the string winding $w$-times around the $\S^1$ fiber 
of the null-brane with radius $R_+$.

We end this subsection
with a comment on the non-zero modes $L_n^{(w)}$ of the
Virasoro generator. $L_n^{(w)}$ does not have a good transformation property
under the ``spectral flow'' \flowalw, since the
integrand of $\h{J}$ is not a conformal current.

\subsec{Torus Amplitute}
As a consistency check of the result in the previous subsection, 
we  compute 
the 1-loop partition sum 
\eqn\Ztr{
Z=\sum_{k,w\in\Z}\Tr_we^{-2\pi ik(\h{J}+p_zR)}
q_-^{L_0-{c\o24}}q_+^{\til{L}_0-{c\o24}},
}
which was computed by the path integral formalism
in \refs{\LMSp,\LMSn,\Mc}.
Here we defined
$q_{\pm}=e^{\mp 2\pi i\tau_{\pm}}$ and $\tau_{\pm}=\tau_1\pm\tau_2$.
$\Tr_w$ means the trace in the $w^{\rm th}$ twisted sector,
and the sum over $k$ imposes the invariance under the orbifold projection.
As emphasized in \LMSp, we have to take the worldsheet metric to be 
Lorentzian since $L_0$ is unbounded from below.

Let us first consider the zero-mode part.
Because the trace is independent of the choice of basis,
we can compute the trace 
in the plane wave basis.
In this basis, we can argue that the trace can be replaced by 
a classical phase space integral which is a Gaussian integral
(see appendix A).\foot{
We can also compute the trace in the $J$-eigenbasis.  
In this case, we have to
redefine the wavefunction $\psi_{\rm LMS}$ in \LMSn\ as
\eqn\newpsi{
\psi_{new}=\exp\lf(-i{J^2\o2p^+x^+}\ri)\psi_{\rm LMS},\quad
\lf[p^2+m^2-{J^2\o (x^+)^2}\ri]\psi_{new}=0,
} 
since $\psi_{\rm LMS}$ is singular in the limit $p^+\riya0$.
Note that $J^2/(x^+)^2$ is the rotation energy around the $\S^1$ fiber.
}
Since $p^-$ appears linearly in $L_0^{(w)}$ in the form $p^+p^-$,
the integral over $p^-$ gives the delta-function $\cob(p^+)$. Therefore,
only $p^+=0$ states contribute to the partition sum. In this computation,
it is important to take $\tau_{\pm}$ to be real.
After performing the Gaussian integral over $p_x$ and $p_z$,
we get the factor
\eqn\clac{
\sum_{k,w\in\Z}\exp\lf(i{\pi R_+^2\o\al'\tau_2}(w\tau_+-k)(w\tau_--k)\ri),
}
which agrees with the contribution of the classical action
in the path integral formalism.

For the non-zero modes, one can show that 
the trace is independent of $w$ by the same reason 
as the path integral computation.
Since $E$ in \JzeroE\ commutes with $N^{(0)}$, we can focus
on a fixed level. By noting that the number basis of oscillators
\eqn\Nbasis{
|N_m^+,N_m^x,N_m^-\ket={1\o\rt{N_m^+!N_m^x!N_m^-!}}
\lf({\al_{-m}^{(0)+}\o\rt{m}}\ri)^{N_m^+}
\lf({\al_{-m}^{(0)x}\o\rt{m}}\ri)^{N_m^x}
\lf({\al_{-m}^{(0)-}\o\rt{m}}\ri)^{N_m^-}|0\ket
}
transforms under the null-boost as
\eqn\trfNbasis{\eqalign{
&e^{ivE}|N_m^+,N_m^x,N_m^-\ket\cr
=&{1\o\rt{N_m^+!N_m^x!N_m^-!}}
\lf({\al_{-m}^{(0)+}\o\rt{m}}\ri)^{N_m^+}
\lf({\al_{-m}^{(0)x}+v\al_{-m}^{(0)+}\o\rt{m}}\ri)^{N_m^x}
\lf({\al_{-m}^{(0)-}+v\al_{-m}^{(0)x}
+{v^2\o2}\al_{-m}^{(0)+}\o\rt{m}}\ri)^{N_m^-}|0\ket,
}}
and using the definition of the conjugate 
$(\al_n^{(0)x})^{\dag}=\al_{-n}^{(0)x},
(\al_n^{(0)\pm})^{\dag}=-\al_{-n}^{(0)\mp}$,
one can easily see that
the matrix element of $q^{N^{(0)}}e^{ivE}$ is $v$-independent:
\eqn\windep{
\bra N_m^+,N_m^x,N_m^-|q^{N^{(0)}}e^{ivE}|N_m^+,N_m^x,N_m^-\ket
=\bra N_m^+,N_m^x,N_m^-|q^{N^{(0)}}|N_m^+,N_m^x,N_m^-\ket.
}
This independence of the twist $w$ in the non-zero mode part is related to 
the fact that in the Fock space there is no eigenstate of $E$ 
with non-zero eigenvalue,
since $\J$ is nilpotent in 
all finite dimensional representations.
Only $J_0$ is diagonalizable on the infinite dimensional space of functions.

Finally, we find that 
the operator formalism 
\Ztr\ gives the same result as the path integral formalism: 
\eqn\Zform{
Z=
{Z^{ghost}Z^{\perp}\o2\pi i\tau_2^2\eta(q_-)^4\eta(q_+)^4}
\int{d^3xdz\o(2\pi\rt{\al'})^4}\sum_{k,w\in\Z}
\exp\lf(i{\pi R_+^2\o\al'\tau_2}(w\tau_+-k)(w\tau_--k)\ri).
}

\subsec{Superstring in the NSR Formalism}
In this subsection, we briefly comment on the extension
to the superstring in the NSR formalism.
In the $w^{\rm th}$ twisted sector,
the mode expansion of the fermions in the $\R^{1,2}$ part
is given by
\eqn\modepsi{
\psi=\sum_{r\in\Z+\nu}e^{-i(r+iw\J)u^+}\psi_r,\quad
\til{\psi}=\sum_{r\in\Z+\nu}e^{-i(r-iw\J)u^-}\til{\psi}_r,
}
where $\nu=\hf$ for the NS sector and $\nu=0$ for the R sector. 
The commutation relation is the same as the untwisted sector:
\eqn\compsi{
\{\psi^\mu_r,\psi^\nu_s\}=\{\til{\psi}^\mu_r,\til{\psi}^\nu_s\}
=\eta^{\mu\nu}\cob_{r+s,0}.
}

The $L_0$ in the twisted sector computed from the stress tensor
\eqn\Tpsi{
T={i\o2}\psi^T\eta\del_+\psi,\quad
\b{T}={i\o2}\til{\psi}^T\eta\del_-\til{\psi}
}
has a similar ``spectral flowed'' form 
as that of the bosonic counterpart:
\eqn\Lzero{
L_0^{(w)}=N_{\psi}+wE_{\psi},\quad
\til{L}_0^{(w)}=\til{N}_{\psi}-w\til{E}_{\psi}
}
where $N_{\psi}$ and $E_{\psi}$ are the number operator
and the null-boost generator, respectively
\eqn\LzeroE{
N_{\psi}=\hf\sum_{r\in\Z+\nu}r\psi_{-r}^T\eta\psi_r,\quad
E_{\psi}={i\o2}\sum_{r\in\Z+\nu}\psi_{-r}^T\eta\J\psi_r.
}
$\til{N}_{\psi}$ and $\til{E}_{\psi}$ are obtained by
putting a tilde on $\psi$ in the above expressions.
The mode expansion and the commutation relation of $\psi^z$ are the same
as the untwisted sector.

The torus amplitude in the superstring case is obtained
by the same formula \Zform\
by including the contribution of
fermions in $Z_{\perp}$ \LMSp.
The important point is that the partition function
of fermions is independent of the twist $w$
\eqn\Zwind{
Z_{\psi}^{(w)}=Z_{\psi}^{(0)},
}
by the same reason as the non-zero modes of the bosonic part.

\newsec{Boundary States on the Null-Brane}
In this section, we construct the boundary states
describing BPS D3- and D2-branes on the null-brane
background which are extended along the $X^\mu$ directions.
The supersymmetry of branes in the null-brane geometry
was considered in \refs{\FSone,\FStwo}. 
It turns out to be easy to construct these states since the boundary condition
is invariant under the null-boost.

\subsec{D3-Branes}
Let us first construct the D3-brane which wraps the whole null-brane.
On the covering space $\R^{1,2}\times\R$,
$X^\mu$ and $Z$ satisfy the Neumann boundary condition.
\eqn\BCDth{
\del_tX^\mu(t=0)|D3\ket=\del_tZ(t=0)|D3\ket=0.
}
In the $w^{\rm th}$ twisted sector
the boundary conditions for modes read
\eqn\DthmBC{\eqalign{
p^\mu|D3\ket_w=p_z|D3\ket_w=0,\quad
(\al_n^{(w)\mu}+\til{\al}_{-n}^{(w)\mu})|D3\ket_w=
(\al_n^{z}+\til{\al}_{-n}^{z})|D3\ket_w=0.
}}
By the spectral flow, the condition for $\al_n^{(w)}$ and 
$\til{\al}_n^{(w)}$ is rewritten as
\eqn\specBC{
(\al_n^{(0)\mu}+\til{\al}_{-n}^{(0)\mu})|D3\ket_w=0.
}
Since this is the ordinary Neumann boundary condition,
we can easily write down the boundary state in the twisted sector: 
\eqn\Bw{\eqalign{
|D3\ket_w&=\exp\lf(\sum_{n>0}-{1\o n}\al_{-n}^{(0)T}\eta
\til{\al}_{-n}^{(0)}-{1\o n}\al_{-n}^z\til{\al}_{-n}^z\ri)|p^\mu=p_z=0\ket_w\cr
&=\exp\lf(\sum_{n>0}-\al_{-n}^{(w)T}\eta{1\o n+iw\J}
\til{\al}_{-n}^{(w)}-{1\o n}\al_{-n}^z\til{\al}_{-n}^z\ri)|p^\mu=p_z=0\ket_w.
}}
This state is invariant under the orbifold action
\eqn\invB{
e^{2\pi i(\h{J}+p_zR)}|D3\ket_w=0,
}
and hence becomes a well-defined state on the orbifold.
This invariance automatically follows from  the boundary conditions  \DthmBC.

For the fermions, the boundary state in the twisted sector is
the same as the untwisted sector
\eqn\psiB{
|D3\ket_{\psi,w}=|D3\ket_{\psi,w=0}.
}
Now the complete
boundary state of D3-brane is obtained by summing
all the twisted sector states
\eqn\wilth{
|D3_{\th}\ket=\sum_{w\in\Z}e^{iw\th}|D3\ket_w.
}
Here we introduced a Wilson line $\th$.

To compute the potential between 
two D3-branes, 
let us consider the cylinder amplitude
\eqn\cyl{
{\cal A}_{\rm closed}=\int_0^{\infty}{ds\o s}
\bra D3_{\th}|e^{-\pi s(L_0+\til{L}_0-{c\o12})}|D3_{\th'}\ket
=\int_0^{\infty}{ds\o s}{\cal A}_0(s){\cal A}'(s).
}
Here we factored out 
the contribution of the zero-mode of $X^\mu$ and $Z$ as ${\cal A}_0(s)$.
${\cal A}'(s)$ is the contribution 
from the transverse coordinates, the fermions,
and the non-zero modes of $X^\mu$ and $Z$.
Plugging the expression of $L_0$ and $\til{L}_0$ \Lzeroflow\ into
this amplitude ${\cal A}_{\rm closed}$,
the contribution of the zero-mode turns out to be
\eqn\Azero{
{\cal A}_0(s)=\int d^3xdz
\sum_{w\in\Z}\exp\lf[-\pi s {w^2R_+^2\o2\al'}-iw(\th-\th')\ri].
}
One can show that the rest of the contribution ${\cal A}'(s)$
is independent of the twist $w$ in the same way as the
computation of the torus amplitude. Especially,
${\cal A}'(s)$ vanishes in the Type IIB string if we 
take the boundary state of fermions 
$|D3\ket_{\psi,w}=|D3\ket_{\psi,w=0}$ 
as that of the BPS D3-brane on the flat space.
Therefore, the effect of the orbifolding is contained
entirely in  ${\cal A}_0(s)$.
By performing the Poisson resummation
\eqn\resumAz{
\sum_{w\in\Z}\exp\lf[-\pi s w^2{R_+^2\o2\al'}-iw(\th-\th')\ri]
={\rt{2\al'}\o\rt{s}R_+}\sum_{n\in\Z}
\exp\lf[-{2\pi\al'\o sR_+^2}\Big(n+{\th-\th'\o2\pi}\Big)^2\ri],
}
we can see that the effective tension of the D3-brane
near $x^+=0$ behaves as
\eqn\etens{
{\cal T}_{D3}\sim \Big[(x^+)^2+R^2\Big]^{-\qu}.
}
This can be thought of as an open string analog of
the time-dependent cosmological constant obtained from
the torus amplitude of the closed string \refs{\LMSp,\LMSn,\Mc}.
Strictly speaking, this is not the tension of the brane defined as 
the strength of the coupling to the graviton.  The time dependence
\etens\ comes from the exchange of twisted closed string states.

We can also compute this cylinder amplitude as a 1-loop amplitude
of the open string.
The open string connecting two D3-branes satisfies the
boundary condition
\eqn\BCopenDth{
\del_\si X^\mu=\del_\si Z=\del_t X_{\perp}=0~~~{\rm at}~~\si=0,\pi,
} 
and the 
Virasoro generator has the standard form $L_0=\al'(p^\mu p_\mu+p_z^2)+N$. 
In the open string channel, the cylinder amplitude \cyl\  is written as
\eqn\opencyl{
{\cal A}_{\rm open}=\int_0^\infty{dt\o t}\sum_{w\in\Z}e^{iw(\th'-\th)}
\Tr \Big[e^{2\pi iw (\h{J}+p_zR)}e^{-2\pi t (L_0-{c\o24})}\Big]
=\int_0^\infty{dt\o t} \til{\cal A}_0(t)\til{\cal A}'(t)
}
In this computation, we have to  perform the
inverse Wick rotation to make the Schwinger
parameter $t$ pure imaginary. 
As above, $\til{\cal A}_0(t)$ is the contribution from
the zero-mode of $X^\mu$ and $Z$, and the rest of the contribution is denoted
as $\til{\cal A}'(t)$. By performing the Gaussian
integral over $p^\mu$ and $p_z$, $\til{\cal A}_0(t)$ is calculated as
\eqn\openzerocnt{
\til{\cal A}_0(t)={i\o 2\pi t^2}\int{d^3x dz\o(2\pi\rt{2\al'})^4}\sum_{w\in\Z}
\exp\lf(-{\pi\o t}{w^2R_+^2\o2\al'}-iw(\th-\th')\ri),
}
and $\til{\cal A}'(t)$ is independent of $w$ by the same reason as above. 
Under the modular transformation $s=1/t$,
${\cal A}_{\rm open}$ agrees
with ${\cal A}_{\rm closed}$ up to a normalization constant.
 

\subsec{D2-Branes}
Next we consider the D2-brane extending along $X^\mu$ and localized
along $Z$ in the covering space.
The boundary condition is given by
\eqn\DtwoBC{
\del_tX^\mu(t=0)|D2\ket=\del_\si Z(t=0)|D2\ket=0.
}
Due to the Dirichlet condition for $Z$, the D2-brane does not
couple to the twisted sector. Therefore,
the boundary state $|D2\ket$ is the same as that on the flat space:
\eqn\Dtwost{
|D2\ket=\sum_{k\in\Z}\exp\lf(\sum_{n>0}-{1\o n}\al_{-n}^T\eta\til{\al}_{-n}
+{1\o n}\al_{-n}^z\til{\al}_{-n}^z\ri)
e^{-iz{k\o R}}\Big|p^\mu=0,p_z={k\o R}\Big\ket_{w=0}.
}
This state is invariant under the orbifold action.
The tension of this brane is time-independent.
It is also easy to construct 
the S-brane localized in all $X^\mu$ directions
since the boundary condition is
null-boost invariant.

\newsec{D1-Branes and Null Scissors}
D1-branes on the null-brane are also BPS states 
when oriented appropriately
on the null-brane. One of the BPS configuration of
D1-brane on the null-brane
is represented by a D1-brane extending along the $x^{\pm}$ directions
on the covering space 
and its infinitely many images under the orbifold group.
The configuration of the original D1-brane and one of its image brane
is called ``null-scissors'' \Bachas.
When the original brane on the covering space is sitting at
$x=a,z=b$, the corresponding brane on the null-brane 
is described by a curve
\eqn\Donecurve{
y={a\o y^+},\quad u=b-{aR\o y^+}.
}
The divergence of $y$ near $y^+=x^+=0$ merely reflects
the coordinate singularity at $x^+=0$.
The phase shift of the $\S^1$ fiber  is finite along the curve \Donecurve 
\eqn\shiftphi{
\lap\phi=\int_{x^+=-\infty}^{x^+=\infty}(dy+A)={\pi a\o R}.
} 
In the case of the parabolic orbifold $R=0$,
this phase shift diverges and the D1-brane winds around the $\S^1$ fiber
infinitely many times near $x^+=0$.
When $R=0$, a D1-brane with $a=0$ and another brane with $a\not=0$
intersect at $x^+=a/2\pi k,~k\in\Z$. 
These intersection points accumulate on
$x^+=0$.

\subsec{Closed String Description}
We can construct the boundary state of D1-brane described above 
as
\eqn\DoonN{
|D1;a,b\ket=\sum_{k\in\Z}e^{2\pi ik(\h{J}+p_zR)}|\h{D1};{a,b}\ket,
}
where  $|\h{D1};{a,b}\ket$ is the 
boundary state on the covering space
\eqn\covDo{
|\h{D1};{a,b}\ket=\exp\lf[\sum_{n>0}{1\o n}(\al_{-n}^+\til{\al}_{-n}^-
+\al_{-n}^-\til{\al}_{-n}^++\al_{-n}^x\til{\al}_{-n}^x
+\al_{-n}^z\til{\al}_{-n}^z)\ri]|p^{\pm}=0,x=a,z=b\ket.
}
The summation over $k$ corresponds to putting infinitely many image branes
under the orbifold action. The state \DoonN\
is invariant under the orbifold action by construction.

The cylinder amplitude between 
$|D1;{a,b}\ket$ and $|D1;{0,0}\ket$ is written as
\eqn\Doneamp{
{\cal A}_{\rm closed}=
\int_0^\infty{ds\o s}
\sum_{k\in\Z}\bra \h{D1};{0,0}|e^{2\pi ik(\h{J}+p_zR)}
e^{-\pi s(L_0+\til{L}_0-{c\o12})}|\h{D1};{a,b}\ket.
}
The contribution from the non-zero modes is $k$-independent as in the previous
section.
The zero-mode contribution is proportional to
\eqn\Doneclen{
\int dx^+dx^-\sum_{k\in\Z}
\exp\lf[-{(2\pi kx^+-a)^2\o2\pi s\al'}-{(2\pi kR-b)^2\o2\pi s\al'}\ri].
}
By performing a Poisson resummation, we can see that the
effective tension becomes time-dependent:
${\cal T}_{D1}\sim [(x^+)^2+R^2]^{-\qu}$.
This time dependence comes from the interactions between infinitely
many image branes on the covering space.
In the next subsection, we will see that the exponent in 
\Doneclen\ can be interpreted as the length of the open string.

\subsec{Open String Description}
The cylinder amplitude \Doneamp\ can be also computed as
an open string 1-loop. 
For a fixed $k$, the open string in question is stretched between
two D1-branes in the null-scissors configuration on
the covering space.
The boundary condition for this open string is
\eqn\origBC{\eqalign{
\si=0:&~~\del_\si X^+=0,\quad X^x=a,\quad \del_\si X^-=0,\quad Z=b\cr
\si=\pi:&~~\del_\si X^+=0,\quad X^x-2\pi k X^+=0,\quad \del_\si 
(X^--2\pi k X^x)=0,\quad Z=2\pi kR.
}}
The mode expansion of $Z$ satisfying this boundary condition is given by
\eqn\Zmodesci{
Z=b+k\lf(R-{b\o2\pi k}\ri)(u^+-u^-)
+i\rt{\al'\o2}\sum_{n\not=0}{1\o n}(e^{-inu^+}-e^{-inu^-})\al_n.
}
To find the mode expansion of $X^\mu$,
it is convenient to perform a T-duality along $X^x$-direction \Bachas. 
The boundary condition in the T-dual picture is given by
\eqn\BCsci{
\del_\si Y|_{\si=0}=0,\quad 
\del_\si Y-\pi v\J\del_tY|_{\si=\pi}=0
}
where $v=2k$,
and the mode expansion is
\eqn\Ysci{
Y=x+i\rt{\al'\o2}\sum_{n\in\Z}{1\o n+iv\J}
\Big(e^{-i(n+iv\J)u^+}+e^{-i(n+iv\J)u^-}-2\cob_{n,0}\Big)\al_n.
}
Taking the T-dual again,
the mode expansion of the original coordinate
is obtained as 
\eqn\origmode{\eqalign{
&X^+=x^++\rt{\al'\o2}(u^++u^-)\al_0^+
+i\rt{\al'\o2}\sum_{n\not=0}{1\o n}(e^{-inu^+}+e^{-inu^-})\al_n^+\cr
&X^x=a+\rt{\al'\o2}\lf[u^+\al_0^x
+{v\o2}(u^+)^2\al_0^++i\sum_{n\not=0}{e^{-inu^+}\o n}
\lf(\al_n^x+v(u^+-{i\o n})\al_n^+\ri)\ri]
-(u^+\lrya u^-) \cr
&X^-=x^-+\rt{\al'\o2}\lf[u^+\al_0^-+{v\o2}(u^+)^2\al_0^x
+{v^3\o6}(u^+)^3\al_0^+\ri] \cr
&~~+i\rt{\al'\o2}\sum_{n\not=0}{e^{-inu+}\o n}\lf[\al_n^-
+v\Big(u^+-{i\o n}\Big)\al_n^x
+v^2\Big({(u^+)^2\o2}-{iu^+\o n}-{1\o n^2}\Big)\al_n^+\ri]
+(u^+\lrya u^-).
}}

From the canonical commutation relation
$[X^\mu(\si),\del_tX^\nu(\si')]=2\pi i\al'\eta^{\mu\nu}\cob(\si,\si')$,
the commutation relation of the non-zero modes is found to be
\eqn\scinzc{
[\al_n^\mu,\al^\nu_m]=(n\eta^{-1}+iv\J\eta^{-1})^{\mu\nu}\cob_{n+m,0}.
}
The zero mode satisfying the boundary condition \origBC\ is given by 
\eqn\alzero{
\al_0^+=\rt{2\al'}p^+,\quad
\al_0^x={v\o\rt{2\al'}}\lf(x^+-{a\o \pi v}\ri),\quad
\al_0^-=\rt{2\al'}p^-,
}
where $[x^{\pm},p^{\mp}]=-i, [p^+,p^-]=[x^+,x^-]=0$.  
Because of the Dirichlet condition of $X^x$ at $\si=0$,
the zero modes $x$ and $p_x$ do not appear in the mode expansion.\foot{
It is well known that the T-dual coordinates satisfying  \BCsci\
are non-commutative at the boundary.
However, the original coordinates \origmode\ are {\it not} non-commutative.
}

As in the case of closed string, 
the commutation relation 
\scinzc\ is written as an ordinary relation using the spectral flow
\eqn\flowsci{
\al_n=S_{v\o n}\al_n^{(0)},\quad
[\al_n^{(0)\mu}\al_m^{(0)\nu}]=n\cob_{n+m,0}\eta^{\mu\nu}.
}
In terms of this spectral flowed oscillator, 
the Virasoro operator of this system is written as\foot{
The zero-mode part of our $L_0^{(k)}$ is different from 
the $L_0$  
of the open string connecting two D2-branes
with a null flux on one of the branes \Bachas. 
This is consistent since
the $X^x$-direction is non-compact.
We just used the T-duality as a solution generating technique
to find the mode expansion in the original D1-brane picture.
In appendix B, we discuss the conpactification of the $X^x$-direction.
}
\eqn\Lzerosci{
L_0^{(k)}=-2\al'p^+p^-+\al'\lf({r\o2\pi\al'}\ri)^2+2kE+N^{(0)}.
}
Here $r^2$ is defined as
\eqn\disr{
r^2=(2\pi kx^+-a)^2+(2\pi kR-b)^2.
}
The $r^2$ term in $L_0^{(k)}$ has a simple physical interpretation.
This term represents the $({\rm mass})^2$ of the string 
stretched between D1-branes at $(x,z)=(2\pi kx^+,2\pi kR)$ and 
$(x,z)=(a,b)$. Although we are not taking the lightcone gauge,
this form of $L_0^{(k)}$ 
suggests that $x^+$ is the good ``time'' coordinate
to describe this system (see fig. 1).
This mass term agrees with the mass of the off-diagonal element
around the diagonal configuration $X^x={\rm diag}(2\pi kx^+,a)$
in the low energy Yang-Mills description of this system \Myers.
It is easy to construct a physical state of the open string
with zero mode only
\eqn\psizero{
\psi=\exp\lf[-ip^+x^-
-i{k^2\o6p^+(\al')^2}\lf(x^+-{a\o2\pi k}\ri)^3
-i{m^2\o2p^+}x^+\ri],~~
m^2=\lf({2\pi kR-b\o2\pi\al'}\ri)^2-{1\o\al'},
}
which is the same as the wave-function of the off-diagonal 
scalar field in the Yang-Mills description \Myers. 
When $p^+=0$, only massless states
are physical, and they are localized at $x^+=a/2\pi k$.

\fig{Two different constant ``time'' slices of null-scissors: 
(a) $x^0$=constant: Two branes are
intersecting at angle $\th=\tan^{-1}(\rt{2}\pi k)$. 
The intersection point moves with the speed of light.
(b) $x^+$=constant: 
Two {\it parallel} branes collide at $x^+=a/2\pi k$ and then
move apart.} 
{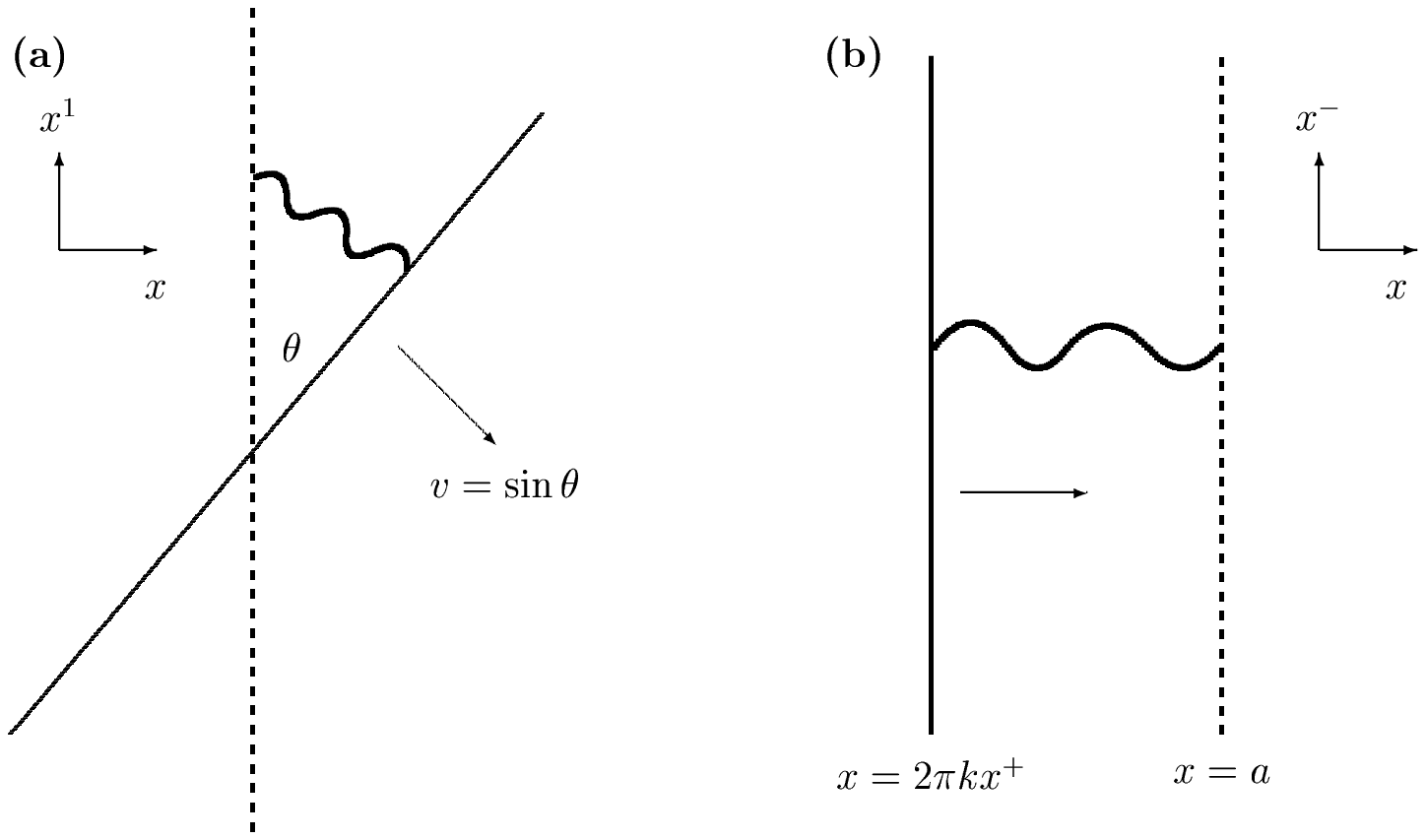}{100mm}\figlabel\parallel

One can easily show that the open string 1-loop amplitude
\eqn\openAsci{
{\cal A}_{\rm open}=\int_0^{\infty}{dt\o t}\sum_{k\in \Z}
\Tr \,e^{-2\pi t(L_0^{(k)}-{c\o24})}
}
agrees with the closed string computation \Doneamp.
Here again, the trace over the non-zero modes is independent
of $k$.

\newsec{Discussions}
In this paper, we have constructed the boundary states
describing D-branes in the null-brane background.
It was pointed out that the low energy theory on a brane in this background
is described by a noncommutative field theory with a time-dependent
noncommutativity \Sethi\ (see also \refs{\Alisha,\Nappi,\Ohta}). 
It is interesting to derive this noncommutativity
from the worldsheet viewpoint.
One can study the worldvolume theory on the brane by
computing a scattering amplitude of open string modes
in a similar way as the closed string computation in \refs{\LMSp,\LMSn,\Mc}.
For instance, the scattering amplitude of open string tachyons
is obtained by a transform of the Veneziano amplitude
in stead of the Virasoro-Shapiro amplitude studied in \refs{\LMSp,\LMSn,\Mc}. 
One can expect that open string amplitudes have a similar
behavior as the closed string ones.
It is also interesting to study the geometry of the null-brane
seen by D-branes.
The instantonic brane, or S-brane, would be a good probe,
and the noncommutative nature of the transverse coordinates of the brane
would have an important role as in \Nekrasov.
The K-theory and the clarification of stable branes in the null-brane
geometry are also interesting.

\vskip 3mm
\centerline{{\bf Acknowledgments}}
I would like to thank B. Craps and S. Sethi for useful discussions,
and C. Bachas for email correspondence.

\appendix{A}{Trace of the Zero-Mode Part}
The trace of the zero-mode part in the twisted sector
typically has the form
\eqn\Zlin{
Z=\Tr \,e^{-iT(\hat{p}^2+\la\hat{x})}
}
where $\h{x}$ and $\h{p}$ are canonically conjugate to each other:
$[\h{x},\h{p}]=i$.
Using the Baker-Campbell-Hausdorff formula
\eqn\BCHf{
\exp X\exp Y=\exp\lf(X+Y+\hf[X,Y]+{1\o12}[X-Y,[X,Y]]+\cdots\ri),
}
we can factorize the operator in the trace as
\eqn\spilitex{
e^{-iT(\hat{p}^2+\la\hat{x})}=
e^{-iT(\hat{p}^2+T\la\hat{p}+{1\o3}T^2\la^2)}e^{-iT\la\hat{x}}.
}
Here, the dots in the
BCH formula vanishes since $[\h{x},[\h{x},\h{p}^2]]=-2$ is a c-number.
Then we can evaluate the trace in the plane-wave basis
\eqn\trel{\eqalign{
Z&=\int dxdp \,\bra p|e^{-iT(\hat{p}^2+T\la\hat{p}+{1\o3}T^2\la^2)}|x\ket
\bra x|e^{-iT\la\hat{x}}|p\ket 
=\int {dx dp\o2\pi} 
e^{-iT(p^2+T\la p+{1\o3}T^2\la^2)}e^{-iT\la x} \cr
&=\int dp \,e^{-iTp^2}\cob(T\la)
=\int {dx dp\o2\pi} e^{-iT(p^2+\la x)}.
}}
Namely, the trace is equal to the classical integral over
the phase space. Notice that it is important to 
use the Lorentzian proper time $T\in\R$.

\appendix{B}{T-dual of the Null-Scissors}
In this appendix, we consider the relation of the null-scissors
and its T-dual configuration. The T-dual of the null-scissors is the
two D2-branes with a constant null flux $F^{-x}$ on one of them.
The open string stretched between these two D2-branes satisfies
the boundary conditions \BCsci, and the mode expansion is given by
\Ysci. The commutators of the oscillators are the same as 
the original D1-brane picture \scinzc, the commutator of the zero modes 
are \Bachas
\eqn\zeroYcom{
[x^\mu,\al_0^\nu]=i\rt{2\al'}\eta^{\mu\nu},\quad
[\al_0^x,\al_0^-]=-iv.
}
These zero-modes $\al_0^\mu$ can be written in terms of the
canonical variables $x^\mu$ and $p^\mu$, obeying
$[x^\mu,p^\nu]=i\eta^{\mu\nu}$, $[x^\mu,x^\nu]=[p^\mu,p^\nu]=0$, as
$\al_0^+=\rt{2\al'}p^+$ and
\eqn\alYgen{
\al_0^x=\rt{2\al'}p^x+fx^++gx,\quad
\al_0^-=\rt{2\al'}p^-+f'x+g'x^+.
}
Here $g$ and $g'$ are arbitrary constants and $f$ and $f'$ are related by
\eqn\fcons{
f+f'={v\o\rt{2\al'}}.
}
The value of these parameters $f,f',g,g'$ can be changed by the
canonical transformation of $x^\mu$ and $p^\mu$.

To see the relation of this T-dual picture with the original 
D1-brane picture, let us compactify the coordinate $Y^x$
on a circle of radius $L$: $Y^x\sim Y^x+2\pi L$.
Then the momentum $p^x$ in this direction is quantized
as $p^x=m/L,~m\in\Z$.
By performing the T-duality along the $Y^x$-direction,
the mode expansion of the original coordinate reads
\eqn\Xtsi{\eqalign{
X^+&=x^++\rt{2\al'}t\al_0^++({\rm oscillators}) \cr
X^x&=x+\rt{2\al'}(\si\al_0^x+v\si t\al_0^+)+({\rm oscillators}).
}}
At the boundary of the open string, these coordinates satisfy
\eqn\BconX{
X^x|_{\si=0}=x,\quad
X^x-\pi v X^+|_{\si=\pi}=x-\pi vx^++\rt{2\al'}\pi \al_0^x.
}
Using the freedom of the choice of parameters $f,f',g,g'$,
we take the following form of the zero-modes:\foot{
The zero-modes with $f'=v/\rt{2\al'}$ and $f=g=g'=0$
are considered in \Bachas.
}
\eqn\alxxp{
\al_0^x=\rt{2\al'}p^x+{1\o\pi\rt{2\al'}}(\pi vx^+-x),\quad
\al_0^-=\rt{2\al'}p^-.
}
Then $X^x-\pi v X^+$ satisfies the Dirichlet boundary
condition at the boundary $\si=\pi$
\eqn\windLtil{
X^x-\pi v X^+|_{\si=\pi}=2\pi\al'p^x=2\pi m\til{L}
}
where $\til{L}=\al'/L$ is the radius of the circle in the $X^x$-direction.
In other words, the ``moving D1-brane'' of the null-scissors
have infinitely many image branes
periodically sitting along the $X^x$-direction.
Now let us take the decompactification limit $\til{L}\riya\infty$.
In the T-dual picture $L\riya0$,
the momentum modes $p^x=m/L$ with $m\not=0$ 
become infinitely massive and hence decouple.
In the original picture, the images of the ``moving brane''
sitting at \windLtil\ with $m\not=0$ go to infinity
and decouple.
Therefore, we can set $p^x=0$ in the decompactification limit,
and we can also set the zero-mode $x$ of $X^x$ to be a constant $a$.
Thus we recover the mode expansion 
\origmode -\alzero\  in the D1-brane picture 
with the non-compact $X^x$ coordinate.
Note that the coordinates $X^\mu$ \origmode\ satisfy
the canonical commutation relation without including the
contribution of the zero-modes $x$ and $p^x$.\foot{
Recall that a boson $\phi=\phi_0+\rt{2\al'}\si\al_0
+\rt{2\al'}\sum_{n\not=0}\al_n/n\,e^{-int}\sin n\si$
obeying the Dirichlet boundary condition 
satisfies the commutation relation 
$[\phi(\si),\del_t\phi(\si')]=2\pi i\al'\cob(\si,\si')$
with the $\cob$-function $\cob(\si,\si')
=1/\pi\sum_{n\not=0}\sin n\si\sin n\si'$. Since the commutator
$[\phi_0,\al_0]$ does not appear in this relation, we can consistently
set $\phi_0$ and $\al_0$ to be c-numbers which specify the
position of D-branes.
}

\listrefs
\bye